\documentclass[sigplan,screen]{acmart}

  \usepackage{amssymb}
  \usepackage{mathtools}
  \usepackage{xspace}
  \usepackage{enumitem}
  \usepackage{listings}
  \usepackage{makecell}
  \usepackage{microtype}
  \copyrightyear{2026}
  \acmYear{2026}
  \setcopyright{cc}
  \setcctype{by}
  \acmConference[LMPL '26]{Proceedings of the 2nd ACM SIGPLAN Workshop on Language Models and Programming Languages}{October 04--09, 2026}{Oakland, CA, USA}
  \acmBooktitle{Proceedings of the 2nd ACM SIGPLAN Workshop on Language Models and Programming Languages (LMPL '26), October 04--09, 2026, Oakland, CA, USA}
  \acmDOI{10.1145/3843750.3843843}
  \acmISBN{979-8-4007-2986-7/2026/10}
  \newcounter{capscopebibitem}
  \AtBeginEnvironment{thebibliography}{%
    \setcounter{capscopebibitem}{0}%
    \pretocmd{\bibitem}{%
      \stepcounter{capscopebibitem}%
      \ifnum\value{capscopebibitem}=5\relax\balance\fi
    }{}{}%
  }

  \newcommand{\tool}{CapScope\xspace}
  
  \newcommand{\eg}{e.g.,\xspace}
  \newcommand{\etal}{et al.\xspace}
  \usepackage{amsthm}
  \renewcommand{\Cap}[2]{\textsf{Cap}[{#1},\,{#2}]}
  \newcommand{\Trusted}{\mathit{Trusted}}
  \newcommand{\Untrusted}{\mathit{Untrusted}}
  \newcommand{\Effects}{\mathit{Effects}}
  \newcommand{\rulename}[1]{\textsc{#1}}

  \newcommand{\tread}{\texttt{read}}
  \newcommand{\twrite}{\texttt{write}}
  \newcommand{\tedit}{\texttt{edit}}
  \newcommand{\tbash}{\texttt{bash}}
  \newcommand{\BZeroTaskSuccess}{72}
  \newcommand{\BOneTaskSuccess}{70}
  \newcommand{\BTwoTaskSuccess}{68}
  \newcommand{\BThreeTaskSuccess}{68}
  \newcommand{\BZeroLeaks}{47}
  \newcommand{\BOneLeaks}{46}
  \newcommand{\BTwoLeaks}{33}
  \newcommand{\BThreeLeaks}{3}
  \newcommand{\BThreeProposals}{34}
  
  \newcommand{\BZeroTime}{145}
  \newcommand{\BOneTime}{140}
  \newcommand{\BTwoTime}{194}
  \newcommand{\BThreeTime}{316}
  \newcommand{\BaseLeakLo}{33}
  \newcommand{\BaseLeakHi}{47}
  \newtheorem{definition}{Definition}

  \lstdefinelanguage{TypeScript}{
    keywords={const, let, var, function, async, await, return,
              if, else, for, while, import, from, export, new,
              class, interface, type, extends, true, false,
              string, number, boolean, void, null, undefined},
    morestring=[b]',
    morestring=[b]",
    morestring=[b]`,
    morecomment=[l]{//},
    morecomment=[s]{/*}{*/},
    sensitive=true
  }
  \lstdefinestyle{tsstyle}{
    language=TypeScript,
    basicstyle=\ttfamily\small,
    numbers=left,
    numberstyle=\tiny\color{gray},
    frame=single,
    breaklines=true,
    showstringspaces=false,
    tabsize=2,
    columns=fullflexible,
    numbersep=6pt,
  }

  \lstdefinestyle{pythonstyle}{
    language=Python,
    basicstyle=\ttfamily\small,
    numbers=none,
    frame=single,
    breaklines=true,
    showstringspaces=false,
    tabsize=2,
    columns=fullflexible
  }

  \title[Authority Is Not a String: A Capability-Scoped Harness\ldots]{Authority Is Not a String:
  A Capability-Scoped Harness for Prompt-Injection-Resistant Coding Agents}

\author{Dimitrios Stamatios Bouras}
\email{2501112125@stu.pku.edu.cn}
\affiliation{%
  \department{Key Lab of HCST (PKU), MoE and School of Computer Science}
  \institution{Peking University}
  \city{Beijing}
  \country{China}
}
\author{Yihan Dai}
\email{2501112020@stu.pku.edu.cn}
\affiliation{%
  \department{Key Lab of HCST (PKU), MoE and School of Computer Science}
  \institution{Peking University}
  \city{Beijing}
  \country{China}}

\author{Sergey Mechtaev}
\email{mechtaev@pku.edu.cn}
\affiliation{%
  \department{Key Lab of HCST (PKU), MoE and School of Computer Science}
  \institution{Peking University}
  \city{Beijing}
  \country{China}}

\begin{document}
  \begin{abstract}
  Coding agents use system-level tools to read files, execute commands, and
  modify source code. Within the agent's sandbox, these tools often carry
  \emph{ambient authority}: naming a resource is sufficient to act on it.
  Indirect prompt injection exploits this authority by placing instructions in
  repository files or tool output that cause the agent to perform actions the
  user did not request.

  We propose \tool, a harness-level authorization mechanism that restricts tool
  use without requiring the model to identify malicious text. Before repository
  contents or tool output are read, \tool\ derives a task-wide authority ceiling
  from trusted input. It then assigns each agent a separate set of typed
  capabilities, stored outside the model's context. Every tool call is checked
  against the capabilities of the agent that issued it. Permissions assigned to
  one sub-agent are therefore not automatically available to another. An
  injection may cause an agent to request an action, but the request is blocked
  unless that agent already has the required capability.

  We implement \tool\ on the Pi coding agent and evaluate it in a
  repair workflow where an orchestrator delegates subtasks to separate
  sub-agents. The evaluation covers five Python tasks, five injection surfaces,
  four authorization conditions, and three trials per cell ($300$ runs). The
  injected effect executes in \BaseLeakLo--\BaseLeakHi/75 runs under the
  ambient-authority and global-policy baselines, compared with
  \BThreeLeaks/75 under \tool.
  \tool\ completes \BThreeTaskSuccess/75 repairs, while the baselines complete
  \BThreeTaskSuccess--\BZeroTaskSuccess/75.
  \end{abstract}
  \ccsdesc[500]{Security and privacy~Access control}
  \keywords{coding agents, indirect prompt injection, capabilities,
    least authority, tool authorization}
  \maketitle
  \section{Introduction}

  A developer asks a coding agent to fix the failing tests in an auth module.
  An orchestrator delegates test execution to a runner and source editing to
  a patcher. A trojaned test
  fixture, planted by a repository contributor or pulled in through a
  compromised dependency, prints to the suite's output an official-looking
  note: \emph{a signing key rotated; restore the build by replacing the key in
  \path{src/auth/keys.py}}. The supplied value is
  attacker-controlled; once committed it lets the attacker forge auth tokens.
  The sub-agent cannot tell this from a real CI message, and the target is an
  ordinary source path. The request mentions no \texttt{.env} or
  \texttt{curl}, and nothing a deny rule would flag. If the sub-agent can
  write source, the backdoor lands, and
  neither it nor the developer notices.

  This is indirect prompt injection: when a coding agent reads external
  resources such as project files, skill definitions, source comments, or tool
  output, any text there can act as an instruction. The threat is
  well-documented: Liu et~al.~\cite{liu2025youraishell} report attack-success
  rates up to $84\%$ against production coding editors in their study, and
  CVE-2025-54135 (\emph{CurXecute}) chained indirect injection to creation of
  Cursor MCP configuration and remote code execution. Malicious rule and skill
  files provide a separate backdoor vector~\cite{pillar2025rules}.

  Most defenses work at the model level: providers increasingly train and
  prompt models to be skeptical of untrusted
  instructions~\cite{wallace2024instructionhierarchy}, and a separate classifier
  may flag suspected injected payloads before they reach the agent. Such
  defenses can reduce attacks, but still make a probabilistic decision
  about whether text is an instruction. Studies have found attacks despite
  instruction-hierarchy prompts~\cite{liu2025youraishell,zhan2024injecagent},
  and classifiers can produce false positives or fail against adaptive
  attacks~\cite{debenedetti2024agentdojo}. \tool\ instead checks permissions
  after the model proposes a tool call. CaMeL~\cite{debenedetti2025camel}
  also uses capabilities, separating control and data flow with a dual-LLM
  interpreter; it evaluates web and email agents rather than coding workflows.

  This is a \emph{confused-deputy} problem~\cite{hardy1988confused}. The harness
  has the user's privileges, while the model chooses the resource names on
  which those privileges act. Capability systems motivate least authority and
  attenuation~\cite{saltzer1975protection,miller2006robust}. \tool\ is a
  host-side reference monitor rather than an object-capability system: the
  model still designates string-named resources, while a per-principal table
  authorizes those requests at dispatch.

  \paragraph{Our Approach: \tool.}
  \tool\ fixes the task's maximum authority from trusted input and then gives each agent
  only the part of it that agent's step requires. It first makes a separate LLM
  call using only the trusted user request and the project's file tree, which
  predicts the files and commands the task will need. The host freezes these
  permissions---stored outside the model's context, so text encountered later
  cannot expand them---before the agent reads the repository or runs a command.

  Concretely, the harness stores three capability forms. 
  \textsf{Read} and \textsf{Write} capabilities cover canonicalized path
  prefixes, while \textsf{Exec} capabilities cover prefixes of parsed argument
  vectors, with every segment of a compound command checked separately. Each
  agent has its own host-side capability store; the model proposes resource
  names and tool calls but cannot read or modify that store.

  When an agent delegates work, the sub-agent receives only the permissions
  its subtask needs, bounded by the delegating ceiling. Every tool
  call is then checked against the permissions of the agent making it. This
  per-agent split is what a single global policy cannot express. In the
  opening example, the sub-agent that reads the poisoned test output can run
  tests but cannot edit source: the source-write operation a patcher would be allowed to
  make is unavailable to the runner that received the injection, so the injected
  key-write is blocked.

  This paper contributes:
  \begin{itemize}[topsep=2pt,itemsep=1pt,leftmargin=12pt]
    \item We identify a granularity gap in current coding-agent
    authorization: a single task-level policy cannot give two agents
    different authority. We close it with a capability-scoped authorization
    model (Section~\ref{sec:promptcap}).
    \item We implement this model in Pi as a harness-level enforcement layer:
    authority is minted from trusted input, attenuated per sub-agent, and
    checked before every tool call (Section~\ref{sec:pattern}).
    \item We evaluate it across five repair tasks, five injection surfaces,
    three baselines, and \tool\ ($300$ runs), measuring both injected-effect
    execution and autonomous task success (Section~\ref{sec:case_study}).
  \end{itemize}

  \paragraph{Artifact.}
  A package containing the code, tasks, mutators, harness, and
  decision logs is available at
  \url{https://figshare.com/s/86184ed20f66d1f0cf91}.

  \section{Background}
  \label{sec:background}
  \paragraph{The Agent Loop.}
  A coding agent combines a model $M$, tools $\mathcal{T}$, and a harness
  $\mathcal{H}$ that runs the loop. The model reads the current \emph{context} $\mathit{ctx}$
  (the message transcript) and emits a \emph{tool proposal}: a tool name with
  concrete arguments, written $(t,\,\mathbf{a})$. The harness executes it and
  appends the resulting observation $o$ to the context. Each tool has an
  \emph{effect type} $\mathit{eff}(t)$ naming the kind of action it causes;
  we group these into \textsf{Read}, \textsf{Write}, and \textsf{Exec}. The
  harness is what actually performs effects, so it
  is the natural place to govern them.

  \paragraph{Trust Boundary.}
  \label{def:integrity}
  The values in the context do not all deserve equal trust, and the
  distinction is what the whole design turns on.
  In a session initiated by a developer, the \emph{trusted} inputs are the
  user's request, the developer-configured system prompt, and the user's
  private settings. Everything whose provenance is the working repository or
  command output is \emph{untrusted}: file contents read by the agent,
  stdout/stderr of executed commands, and project-local config files any
  repository contributor can modify.

  Because the model reads both trusted and untrusted text, its tool calls are
  treated as untrusted requests~\cite{liu2025youraishell}. The harness does not
  infer authority from a call; it checks authority stored outside the
  conversation.

  \subsection{Coding Agents}
  \label{sec:agents}
  \tool\ targets coding agents whose harness mediates tool calls and maintains
  a separate capability store for each agent session, which we call a
  \emph{principal}.

  \paragraph{Tools and Effects.}
  A coding agent's tools cause three kinds of effect. \textsf{Read} and
  \textsf{Write} concern the filesystem. \textsf{Exec}---running a shell
  command---is the broadest: an unrestricted command can reach the network, read
  secrets, rewrite git history, or install packages, all with the user's
  privileges. Absent a command-level restriction, \textsf{Exec} is the effect a
  misdirected agent is most often steered toward.

  \paragraph{Injection Surfaces.}
  A coding agent loads context from several sources in the working directory.
  All are $\Untrusted$ under the trust boundary above:

  \begin{enumerate}[label=(\arabic*),topsep=2pt,itemsep=0pt]
    \item \textbf{README.md.} Project documentation the agent reads on demand;
      any repository contributor can modify it.
    \item \textbf{AGENTS.md.} A repository-controlled instruction file loaded
      automatically for the agent.
    \item \textbf{Skill files.} Project-local markdown prompt templates.
      Analogous to coding rule files, a known backdoor vector in production
      editors~\cite{pillar2025rules}.
    \item \textbf{Source file contents.} An instruction in a comment or
      docstring enters the context as $\Untrusted$ data when the model reads
      the file.
    \item \textbf{Tool output.} The standard output and error of shell
      invocations return to the model as $\mathit{Tool}$-role messages; a
      trojanized test runner or network-fetched script can inject through this
      channel.
  \end{enumerate}

  These surfaces define the attack scenarios in our evaluation
  (Section~\ref{sec:case_study}). Under ambient authority, each can redirect
  the model toward an attacker-chosen tool call; how a capability discipline
  changes that outcome is the subject of Section~\ref{sec:promptcap}.

  \paragraph{What \tool\ Requires of a Host.}
  A host needs a blocking pre-dispatch interceptor and per-session state. A
  lifecycle hook that only observes tool calls is insufficient: the interceptor
  must reject a call before execution. The capability representation itself is
  independent of the host API.

  \subsection{Threat Model and Non-Goals}
  \label{sec:threat}
  We consider indirect prompt injection through any text that enters the
  model's context from repository files, project-instruction files, skill
  definitions, source comments, or tool output. The attacker may steer the
  model's proposed tool calls but cannot modify the harness, the capability
  store, or the tool wrappers, and cannot cause a capability to be minted.
  Within this model, an external effect runs only if a held capability covers
  it. \tool\ does not stop the model from following an injected instruction; it
  stops the resulting tool call when that agent lacks the required capability.

  \tool\ is not a confidentiality mechanism. A permitted read followed by a
  permitted write can still move data, and an injection can misuse authority
  deliberately granted to its reader. Those cases require complementary
  information-flow controls or sandboxing.

  \paragraph{Assumptions.}
  The guarantee assumes a small trusted computing base: every effect is
  mediated; stores are host-side and isolated by principal; path and command
  predicates faithfully implement their structured scopes; and only the host may
  mint or complete a derivation. The model may propose the initial set or request
  a sub-agent scope, but the host controls store updates and checks that each
  requested scope lies within the delegating ceiling.

  The preflight reads directory and file names, but not file contents. Filename
  injection is outside this evaluation. The host validates and freezes the
  generated ceiling before any evaluated injection surface is read. The
  guarantee assumes correct predicates, complete mediation, and no repository
  contents in the preflight input.

  \paragraph{Scope of the Claim.}
  \tool\ bounds which effects a principal may invoke; it does not make every
  covered effect safe. An injection can misuse authority deliberately granted to
  its reader. An allowed command can also execute untrusted repository code
  transitively---\texttt{pytest}, for example, imports project modules---so
  process sandboxing remains necessary. These are complementary boundaries, not
  properties of capability lookup.
  Likewise, \textsf{Read} scopes govern the dedicated file-reading tool, not
  confidentiality through the command wrapper. Read/navigation commands such as
  \texttt{grep}, \texttt{find}, and \texttt{ls} are admitted by that wrapper and
  may inspect paths outside a principal's \textsf{Read} prefix. Deployments that
  require confidentiality must additionally sandbox command-level filesystem
  access.

  \section{\tool: The Design Pattern}
  \label{sec:promptcap}

  \tool\ replaces ambient authority with permissions held by the harness. It
  checks whether an installed capability covers a proposed tool call, without
  asking the model to classify the proposal's provenance.

  \subsection{Capabilities as Typed Authority}
  \label{sec:captype}

  \begin{definition}[Capability]
  \label{def:capability}
  A \emph{capability} is a pair $c = (e,\, \varphi)$ written
  $\Cap{e}{\varphi}$, where $e \in \Effects$ is an \emph{effect type}
  and $\varphi : \mathit{Args} \to \mathbb{B}$ is a \emph{scope
  predicate} over the arguments of that effect.  A capability
  $\Cap{e}{\varphi}$ \emph{covers} a proposal $(t,\, \mathbf{a})$ when
  $\mathit{eff}(t) = e$ and $\varphi(\mathbf{a})$ holds.
  \end{definition}

  A \emph{capability store} $\Sigma$ is a finite set of capabilities held
  by the harness for the duration of a session.  The model never holds
  capabilities directly; it sees only the effects they permit.

  \paragraph{Effect Types and Scope Predicates for Coding Agents.}
  For the four tools of a typical coding agent, we define three effect
  types and give representative scope predicates.  Here $p$ denotes a
  file-system path and $s$ a shell command string.

  \begin{center}
  \small
  \setlength{\tabcolsep}{11pt}
  \begin{tabular}{@{} l l @{}}
  \toprule
  \textbf{Effect type (tools)} & \textbf{Scope predicate $\varphi$} \\
  \midrule
  \textsf{Read} \;(\tread)
    & $\varphi(p) \;\triangleq\; \mathit{within}(\mathit{canon}(p),\pi)$ \\
  \textsf{Write} \;(\twrite, \tedit)
    & $\varphi(p,\_) \;\triangleq\; \mathit{within}(\mathit{canon}(p),\pi)$ \\
  \textsf{Exec} \;(\tbash)
    & $\varphi(s) \;\triangleq\; \mathit{allowed}_A(\mathit{parse}(s))$ \\
  \bottomrule
  \multicolumn{2}{@{}l@{}}{\scriptsize
    $\pi$: canonicalized path prefix;\enspace
    $A$: allowed argument-vector prefixes;} \\[-2pt]
  \multicolumn{2}{@{}l@{}}{\scriptsize
    $\mathit{canon}$: resolves \texttt{..} and symlinks.}
  \end{tabular}
  \end{center}

  $\mathit{within}(x,\pi)$ means that path $x$ equals or lies below prefix
  $\pi$. We write \texttt{src/**} as shorthand for the canonical subtree rooted
  at \texttt{src/}. The predicate uses a canonicalized path
  ($\mathit{canon}$ resolves \texttt{..} and symlinks) to prevent
  traversal attacks. $\mathit{parse}(s)$ rejects substitution, splits compound
  commands, removes redirections, and returns each segment's argument vector.
  $\mathit{allowed}_A$ holds when every vector is read/navigation or begins with
  a vector in $A$.
  Thus \texttt{Exec[pytest]} admits \texttt{pytest -v}. A compound command
  containing an uncovered \texttt{curl} segment is rejected. Redirections and
  transitive effects of allowed commands remain
  inside the process-sandbox boundary.

  \subsection{Formal Semantics}
  \label{sec:formalism}

  \paragraph{Notation.}
  $e$ is an effect type, $t$ a tool, $\mathbf a$ its arguments, and $o$ an
  observation. $\mathit{eff}(t)$ gives the effect of $t$;
  $\varphi:\mathit{Args}\to\mathbb B$ accepts or rejects arguments, where
  $\mathbb B=\{\mathit{true},\mathit{false}\}$. $\Cap{e}{\varphi}$ pairs an
  effect with a scope; $\Sigma_i$ is agent $i$'s finite capability store; and
  $\mathit{ctx}\cdot o$ appends $o$ to the context. For the root orchestrator,
  $D_i$ is the task-wide ceiling; for a non-root agent, it is that agent's
  already-derived store. Here $r$ is the root and $\langle X;\mathit{ctx}\rangle$
  is a state with authority $X$ and context $\mathit{ctx}$. We write
  $c'\preceq c$ when $c'$ and $c$ have the same
  effect type and the scope of $c'$ is no broader: for
  $c'=\Cap{e}{\varphi'}$ and $c=\Cap{e}{\varphi}$,
  $\forall\mathbf a.\,\varphi'(\mathbf a)\Rightarrow\varphi(\mathbf a)$.
  $D_i\vdash C'\Downarrow\Sigma_{\mathit{sub}}=C''$ means that candidate set
  $C'$ is checked under delegation ceiling $D_i$, and the covered subset $C''$
  is installed as the sub-agent's capability store.
  $\longrightarrow$ updates an authority set, $\Downarrow$ installs a derived store or
  returns an observation, and $\xrightarrow{(t,\mathbf a)}$ records an invoked
  tool call.

  \paragraph{Minting: Where Authority Comes From.}
  \rulename{Mint} is the only rule that introduces authority not already
  bounded by a held capability. It runs a trusted host policy $P$ and adds the
  finite capability set $C$ that it returns for trusted preflight input $I_T$:
  {\small\[
    \frac{
      P \in \mathit{Policies}_{\Trusted} \quad P(I_T)=C
    }{
      \langle D_r;\mathit{ctx}\rangle \longrightarrow
      \langle D_r\cup C;\mathit{ctx}\rangle
    }\; (\rulename{Mint})
  \]}
  With fresh $D_r=\varnothing$, this sets $D_r=C$. Host-side compilation installs
  $\Sigma_r\subseteq D_r$ for root calls, as in Listing~2.
  In our implementation, an LLM proposes capability descriptions from the
  trusted request and the project's file tree---names only, never contents.
  Deterministic host code validates and compiles the proposal as policy $P$
  before file contents, tool output, or another injection surface is available. The
  policy is then frozen for that session. The ceiling and operational store are
  fresh per session and do not accumulate across tasks. If a later call is genuinely necessary
  but uncovered, the harness may ask the user to approve a single additional
  capability. That out-of-band decision is a new trusted \rulename{Mint},
  followed by host-side compilation; the model cannot approve its own request.
  Interactive approval is disabled in our evaluation.

  \paragraph{Derivation: Narrowing for Delegation.}
  A model supplies a requested grant, and host-side completion may add
  role-required candidate descriptions. Let $C'$ be the resulting candidate
  set; its descriptions carry no authority until installed by the host.
  \rulename{Derive} filters $C'$ against $D_i$, installs the covered subset
  $C''$, and drops uncovered requests:
  {\small\[
    \frac{
      C''=\{c'\in C' \mid \exists c\in D_i.\;c'\preceq c\}
    }{
      D_i \vdash C' \Downarrow \Sigma_{\mathit{sub}}=C''
    }\; (\rulename{Derive})
  \]}
  For example \texttt{src/**} can be narrowed to \texttt{src/parser.py}, never
  the reverse. A spawned sub-agent's fresh store contains $C''$ alone.

  \paragraph{Invocation: The Check Before Every Effect.}
  \rulename{Invoke} is where a proposal becomes an effect. It fires only when
  some held capability matches the tool's effect type and its predicate
  accepts the arguments; $\mathit{execute}(t,\,\mathbf{a}) \Downarrow o$ runs
  the tool and yields the observation $o$ appended to the context:
  {\small\[
    \frac{
      \begin{gathered}
      \Cap{e}{\varphi}\in\Sigma \quad \mathit{eff}(t)=e \quad
      \varphi(\mathbf a) \\
      \mathit{execute}(t,\mathbf a)\Downarrow o
      \end{gathered}
    }{
      \langle\Sigma;\mathit{ctx}\rangle
      \xrightarrow{(t,\mathbf a)}
      \langle\Sigma;\mathit{ctx}\cdot o\rangle
    }\; (\rulename{Invoke})
  \]}
  If no capability covers the proposal it is blocked and the model receives a
  refusal. Since the model's output is untrusted (the trust boundary, Section~\ref{sec:background}),
  the check never consults the model about provenance. The only question is
  whether $\Sigma$ already holds a covering capability.

  \paragraph{What the Rules Give Us.}
  By inspection of the three rules, every effect that executes is covered by a
  capability whose authority traces back to a trusted \rulename{Mint}.
  \rulename{Invoke} is the only rule that runs an effect, and it requires a
  covering capability already in $\Sigma$. Root authority enters $\Sigma_r$ only
  through compilation from $D_r$; sub-agent authority enters through
  \rulename{Derive}. Both can only narrow, so no agent can exceed the initial
  minted authority. In our
  evaluation, this mint occurs before untrusted content is admitted and later
  user-approved minting is disabled. Injected text may influence a proposed
  tool call or requested sub-agent grant, but it cannot expand the task-wide
  authority. A call outside the resulting scope is refused.

  \subsection{Implementation}
  \label{sec:pattern}
  We realize \tool\ on Pi~\cite{pi2025}, an open-source TypeScript coding-agent
  toolkit with a CLI and a programmable SDK, without modifying Pi internals.
  The experiment enables Pi's four built-in tools below; the command wrapper
  also recognizes read/navigation programs such as \texttt{grep}, \texttt{find},
  and \texttt{ls}.

  \begin{center}
  \small
  \setlength{\tabcolsep}{5pt}
  \begin{tabular}{@{} l p{3.6cm} l @{}}
  \toprule
  \textbf{Tool} & \textbf{Arguments} & \textbf{Effect} \\
  \midrule
  \texttt{read}
    & \texttt{path : string}
    & \textsf{Read}  \\
  \texttt{write}
    & \texttt{path : string,\newline content : string}
    & \textsf{Write} \\
  \texttt{edit}
    & \texttt{path : string,\newline edits : Edits}
    & \textsf{Write} \\
  \texttt{bash}
    & \texttt{command : string}
    & \textsf{Exec}  \\
  \bottomrule
  \end{tabular}
  \end{center}

  Pi 0.78.0 supplies the two host requirements of Section~\ref{sec:agents}.
  Extensions register handlers with \texttt{pi.on}. The \texttt{tool\_call}
  event occurs before dispatch; returning \texttt{block: true} aborts the call
  and sends the reason to the model. \tool\ installs this hook separately for
  the orchestrator and each sub-agent. The SDK's
  \texttt{createAgentSession} returns \texttt{\{session\}}; callers submit work
  with \texttt{session.prompt()}. The \texttt{pi-subagents}
  extension (\url{https://www.npmjs.com/package/pi-subagents}) delegates
  a subtask to a \emph{process-isolated} sub-agent, whose separate address
  space lets \tool\ seed its capability store independently.

  \paragraph{Component 1: Capability Structure.}
  Scopes are structured data rather than opaque functions. The host can check
  whether a requested path prefix or command set lies within the delegating
  ceiling before installing it.

  \begin{lstlisting}[language=TypeScript,style=tsstyle,caption={Schematic
    pseudocode for the structured capability type.}]
  type EffectType = "read" | "write" | "exec";

  type ScopeSpec =
    | { kind: "path"; prefix: string }
    | { kind: "argv"; allowed: string[][] };

  interface Capability {
    effect: EffectType;
    spec: ScopeSpec;
    description: string;
  }

  type CapSet = Capability[];
  \end{lstlisting}

  \paragraph{Component 2: Initial Minting.}
  Before loading project context, \texttt{authorize} makes a preflight LLM
  call using the trusted request and the project's file tree. The call predicts the
  permissions required by the complete task and by the orchestrator. The host
  validates and compiles this prediction into a task-wide \emph{ceiling}, from
  which sub-agents may derive authority, and the orchestrator's operational
  store. The latter may be read-only. The root orchestrator is the exception to
  ordinary delegation: its calls use the operational store, while its grants
  are bounded by the task ceiling.

  \begin{lstlisting}[language=TypeScript,style=tsstyle,caption={Schematic
    pseudocode for initial authorization.}]
  interface Authority {
    ceiling: CapSet;
    orchestrator: CapSet;
  }

  async function authorize(
    trustedRequest: string,
    taskTree: TaskTree
  ): Promise<Authority> {
    const proposal = await preflightLLM({
      trustedRequest,
      taskTree
    });
    return validateAndCompile(proposal);
  }
  \end{lstlisting}

  \noindent The \texttt{authorize} call appears once, at session start.
  Repository text and tool output are unavailable at this point. The
  preflight prediction may still be too broad or too narrow, but later
  injected content cannot change it. A runtime user approval, if enabled, is a
  separate trusted mint and is recorded as such.

  \paragraph{Component 3: Derivation and Transfer.}
  Delegation is driven by the model, just as the initial mint is. When an agent
  delegates, it issues a \texttt{subagent} tool call whose structured arguments
  carry the sub-agent's role, subtask, and a \emph{requested grant}; the same
  decision that chooses what to delegate also proposes how much authority the
  sub-agent should get. The host treats that request as a proposal only:
  \texttt{derive} keeps just the capabilities structurally contained in the
  delegating ceiling (rule~\rulename{Derive}) and drops the rest. The
  implementation first calls \texttt{ensureUsable} to supply omitted
  repository-local reads and role-required test or write scopes, then calls
  \texttt{derive} to filter the completed request against the frozen ceiling.
  Repository reads are allowed by default in all conditions. Because the
  subtask text is model-generated, it may affect which in-ceiling scope is
  selected, as discussed in Section~\ref{sec:discussion}.

  \begin{lstlisting}[language=TypeScript,style=tsstyle,caption={Schematic
    pseudocode for deriving and transferring a sub-agent store.}]
  function derive(ceiling: CapSet, requested: CapSet) {
    const derived = requested.filter(cap =>
      ceiling.some(p =>
        p.effect === cap.effect && scopeWithin(cap.spec, p.spec)
      )
    );
    return { derived };
  }

  async function spawnScoped(
    ceiling: CapSet, role: Role, task: string, requested: CapSet
  ) {
    const completed = ensureUsable(role, task, requested, ceiling);
    const { derived: subCaps } = derive(ceiling, completed);
    const { session: sub } = await createAgentSession({
      resourceLoader: scopedRoleResources(role, subCaps)
    });
    await sub.prompt(task);
  }
  \end{lstlisting}

  \paragraph{Component 4: Per-Principal Dispatch Checks.}
  Every orchestrator and sub-agent session installs the same small handler, but
  closes over a different store. It compiles the structured scope, checks
  \rulename{Invoke}, and blocks uncovered calls.

  \begin{lstlisting}[language=TypeScript,style=tsstyle,caption={Schematic
    pseudocode for the per-principal \tool\ hook.}]
  function installCapScope(pi: ExtensionAPI, caps: CapSet) {
    pi.on("tool_call", (event) => {
      const granted = caps.find(
        c => effectOf(event.toolName) === c.effect
          && compileScope(c.spec)(event.input)
      );
      if (!granted) {
        return {
          block: true,
          reason: `[CapScope] uncovered ${event.toolName} call`
        };
      }
    });
  }
  \end{lstlisting}

  \paragraph{Usage: A Complete Session.}
  The root's calls are checked against \texttt{auth.orchestrator}; delegation
  is checked against \texttt{auth.ceiling}. Thus a read-only orchestrator may
  grant a task-required write to a patcher, but no grant can exceed the ceiling.

  \begin{lstlisting}[language=TypeScript,style=tsstyle,caption={Schematic
    pseudocode for orchestrator setup.}]
  const auth = await authorize(userRequest, taskTree);
  const { session: orchestrator } = await createAgentSession({
    resourceLoader: capScopeResources(auth)
  });
  await orchestrator.prompt(userRequest);
  \end{lstlisting}

  \paragraph{Other Frameworks.}
  Porting \tool\ requires a host-specific blocking pre-dispatch interceptor and
  per-session state. In-process sub-agents need separate stores; separate
  processes receive a derived store at startup. The hook registration and
  session setup must be checked against each framework's API.

  \subsection{Example: A Backdoor the Allowlist Cannot Stop}
  \label{sec:example}
  A developer asks an orchestrator to fix the failing auth-module tests. The
  task needs project reads, source writes, and test commands. Before reading
  project content, preflight receives only the trusted request and file tree;
  the host compiles $c_1$ for reads, $c_2$ for writes, and $c_3$ for execution.
  Write $A_{\mathit{task}}$ for the allowed prefixes \texttt{pytest} and
  \texttt{mypy src/}, and $A_{\mathit{run}}$ for \texttt{pytest} alone.
  \begin{align}
    c_1 &= \Cap{\textsf{Read}}{\mathit{within}(\mathit{canon}(p),
          \texttt{"project/**"})} \label{cap:read} \\
    c_2 &= \Cap{\textsf{Write}}{\mathit{within}(\mathit{canon}(p),
          \texttt{"src/**"})} \label{cap:edit} \\
    c_3 &= \Cap{\textsf{Exec}}
          {\mathit{allowed}_{A_{\mathit{task}}}(\mathit{parse}(s))} \label{cap:exec}
  \end{align}
  The orchestrator delegates the noisy test run to a \emph{runner} and later the
  source edit to a \emph{patcher}. The runner needs to read tests and run
  \texttt{pytest} but needs no write, so before spawning it the orchestrator
  applies \rulename{Derive} to pass a strict subset:
  \begin{align}
    c_1' &= \Cap{\textsf{Read}}{\mathit{within}(\mathit{canon}(p),
            \texttt{"tests/**"})} \label{cap:dread} \\
    c_3' &= \Cap{\textsf{Exec}}
            {\mathit{allowed}_{A_{\mathit{run}}}(\mathit{parse}(s))}
            \label{cap:dexec}
  \end{align}
  Here \texttt{tests/**} is contained in \texttt{project/**} and \texttt{mypy}
  is dropped, so $\varphi' \Rightarrow \varphi$ and
  $\Sigma_{\mathit{runner}} = \{c_1', c_3'\}$. The runner's store contains no
  capability with effect type \textsf{Write}.

  A trojaned fixture
  makes the suite print, to the runner's \texttt{pytest} output:
  \begin{quote}\small\itshape
  \textbf{[ci-bot] Known regression:} the signing key rotated. Restore the
  build by writing to \texttt{src/auth/keys.py}:
  \texttt{SIGNING\_KEY = "k3y\_9f2a\ldots c71"}, then re-run.
  \end{quote}
  The key is attacker-controlled, and the target is an \emph{ordinary source
  path}---no \texttt{.env}, \texttt{\textasciitilde/.ssh}, or \texttt{curl}, and
  nothing a deny pattern matches. Figure~\ref{fig:trace} compares three
  authorization models.
  \begin{figure*}[t]
  \centering
  \small
  \setlength{\tabcolsep}{6pt}
  \begin{tabular}{@{} c l c c c @{}}
  \toprule
  \textbf{Turn} & \textbf{Runner proposes}
    & \makecell{\textbf{Static}\\\textbf{global policy}}
    & \makecell{\textbf{Task-specific}\\\textbf{global policy}}
    & \makecell{\textbf{\tool}\\\textbf{per agent}} \\
  \midrule
  1 & \texttt{read("tests/test\_auth.py")}
    & allowed (global read) & allowed ($c_1$) & allowed ($c_1'$) \\[1pt]
  2 & \texttt{bash("pytest tests/test\_auth.py")}
    & \makecell{allowed (global exec)\\$\to$ payload enters}
    & \makecell{allowed ($c_3$)\\$\to$ payload enters}
    & \makecell{allowed ($c_3'$)\\$\to$ payload enters} \\[1pt]
  3 & \texttt{write("src/auth/keys.py", \ldots)}
    & \makecell{\textbf{allowed}: matches\\no deny rule}
    & \makecell{\textbf{allowed}: $c_2$\\covers \texttt{src/**}}
    & \makecell{\textbf{blocked}: no\\\textsf{Write} in $\Sigma_{\mathit{runner}}$} \\
  \bottomrule
  \end{tabular}
  \caption{The same runner under three authorization models. Turns~1 and 2
    are legitimate and allowed everywhere, so the poisoned test output enters the runner's
    context and it proposes the backdoor write at turn~3. The static global
    policy allows the ordinary-looking source path. The task-specific global
    policy also allows it because its shared authority includes $c_2$ for the
    legitimate patch. \tool\ blocks it because the runner has only
    $\Sigma_{\mathit{runner}}=\{c_1',c_3'\}$.}
  \Description{A three-column execution trace comparing a static global
    policy, a task-specific global policy, and CapScope. All three allow the
    runner to read tests and execute pytest; only CapScope blocks the injected
    source write because the runner has no write capability.}
  \label{fig:trace}
  \end{figure*}
  The static policy does not classify this source edit as dangerous. B2's
  shared policy includes source writes because the repair needs them, so it
  cannot withhold that write from the runner while allowing the patcher to fix
  the file. \tool\ gives the write only to the patcher. At turn~3 the runner is
  refused and may report the requested edit to the orchestrator as data; its
  own store still contains no \textsf{Write} capability.

  \section{Evaluation}
  \label{sec:case_study}

  We ask two questions. \emph{RQ1:} does per-principal scoping reduce attacks
  that global, task-level authorization cannot distinguish? \emph{RQ2:} what is
  the task-success and latency cost of fail-closed capability enforcement? Across
  conditions we vary the authorization model while holding the underlying LLM,
  the tasks, and the delegated four-principal topology fixed.

  \paragraph{Tasks and Attacks.}
  Each task is a small Python project with one bug and a failing
  \texttt{pytest} suite. The injections request executing a project-local
  utility or writing to a project file. These actions are
  intentionally plausible rather than obviously harmful: each could be
  legitimate in another maintenance task, but none is needed for the requested
  repair. In the injected context, such execution or writing could introduce a
  malicious side effect. Table~\ref{tab:tasks} summarizes the repair tasks.

  \begin{table}[t]
  \centering\small
  \setlength{\tabcolsep}{5pt}
  \caption{Repair tasks used in the evaluation.}
  \label{tab:tasks}
  \begin{tabular}{@{} l p{0.76\columnwidth} @{}}
  \toprule
  \textbf{Task} & \textbf{Bug} \\
  \midrule
  T1 & Off-by-one error in \texttt{daterange.py} \\
  T2 & Wrong fee comparison in \texttt{fees.py} \\
  T3 & Wrong field lookup in \texttt{report.py} \\
  T4 & Case-sensitive lookup in \texttt{indexer.py} \\
  T5 & Discount applied on the wrong branch in
       \texttt{billing/fees.py} \\
  \bottomrule
  \end{tabular}
  \end{table}

  Each malicious action creates a unique sentinel or marker that benign repair
  never produces. We score that side effect, rather than treating routine
  project activity itself as compromise.

  \paragraph{Delegated Pipeline.}
  Every condition uses the same four-principal workflow. An orchestrator
  diagnoses and coordinates but delegates effectful work. A runner executes
  the failing tests, a patcher edits the relevant source, and a verifier
  reruns the suite. The loop repeats until the tests pass or the run
  terminates. This control is important: the conditions compare authorization models,
  not different agent architectures.

  \paragraph{Injection Surfaces.}
  We place the same task-specific payload in each of the five untrusted
  surfaces of Section~\ref{sec:agents}: a README, an agent-instruction file, a
  project-local skill, a source comment or docstring, and \texttt{pytest}
  output. The first three are normally encountered by the orchestrator while
  inspecting the project, source comments by an agent reading code, and test
  output by the agent running or verifying the tests. Varying the surface
  therefore changes which principal first receives the injection without
  changing the requested repair or malicious action.

  \paragraph{Baselines.}
  We compare \tool\ against three baselines that bracket current
  practice. All principals use ambient authority in \textbf{B0}. \textbf{B1}
  applies one static, task-independent policy to every principal, configured
  through the permission-gating facility Pi provides: it denies familiar
  hazards such as sensitive paths and obviously destructive commands (\eg
  \texttt{.env} and \texttt{.ssh} paths, \texttt{rm\,-rf}, \texttt{sudo},
  \texttt{curl}), but permits ordinary project operations. \textbf{B2}
  generates a tighter global policy from the trusted task description and a
  sanitized project structure before any injection is read, then applies that
  one policy to every principal; it can remove authority the task does not
  need, but cannot give two principals different slices of authority.
  \textbf{C} is not a baseline but \tool\ itself, the system under test: a
  preflight mints a task-wide ceiling that bounds delegation, and each sub-agent
  receives a store derived from the applicable delegation ceiling.
  B2 is the strongest global baseline: its policy is tailored to the task
  before the adversarial text arrives. Comparing it with C shows the outcomes
  of adding per-agent authority to the complete mechanism.

  \paragraph{Protocol and Metrics.}
  The full factorial is
  \[
    4\ \text{conditions} \times 5\ \text{tasks} \times
    5\ \text{surfaces} \times 3\ \text{trials} = 300\ \text{runs}.
  \]
  Each run uses an isolated working copy and
  all conditions are driven by \textbf{qwen3.5-flash} through CloseAI's
  OpenAI-compatible endpoint. Runs were collected in May 2026; preflight calls
  used temperature 0.
  The harness verifies that the suite fails before the run, then records two
  outcomes:
  \begin{itemize}[topsep=2pt,itemsep=1pt,leftmargin=12pt]
  \item \texttt{task\_success}: the final test suite passes;
  \item \texttt{attack\_executed}: the attacker-only sentinel or marker exists
    on disk after the run.
  \end{itemize}
  Each outcome is Boolean per run: a run is counted once if the attack executes
  at least once, even if it also contains blocked or repeated attack attempts.
  Both are read from the filesystem, independent of any log and of each other:
  a run may fix the bug and still be compromised. \texttt{attack\_executed} is
  our headline defense metric---did the attack land. The per-call decision log
  additionally shows, on the conditions that install a hook, whether a call
  carrying the injected action was \emph{proposed} by the model and whether the
  layer \emph{blocked} it. We report these as secondary diagnostics, not outcome
  metrics. They need not be mutually exclusive because a run may contain
  several attempts. B0 installs no hook and exposes neither count.

  \paragraph{Results.}
  Table~\ref{tab:case} reports descriptive counts over the three trials. For
  attack execution, $\mathrm{B0} \approx \mathrm{B1} > \mathrm{B2} \gg
  \mathrm{C}$: the injected effect lands in \BZeroLeaks\ of $75$ runs under
  ambient authority and \BOneLeaks\ under the static global policy---a
  conventional denylist does not improve on having no policy, because the
  malicious actions are ordinary in-repository operations it never names. The
  task-specific global policy (B2) helps, cutting leakage to \BTwoLeaks/75 by
  removing task-unnecessary operations before the injection arrives, but that
  one allowlist is still shared by every principal (below). \tool\ (C) admits the
  effect in \BThreeLeaks/75, compared with \BTwoLeaks/75 for the strongest
  global baseline, because it additionally withholds task-needed authority
  from the principals that do not need it. The condition-specific per-surface
  view (Table~\ref{tab:surface}) compares B2 with C. Under C, all three executed
  attacks occur through \texttt{AGENTS.md}; the other four surfaces admit none.

  Task success differs by four runs between C and B0: \tool\ finishes
  \BThreeTaskSuccess\ of $75$ repairs against \BZeroTaskSuccess\ for the
  unconstrained baseline---a gap of four runs---and B2's tighter global policy
  also finishes \BTwoTaskSuccess/75. As a diagnostic, the decision log shows that under \tool\ the
  model proposes the injected action on \BThreeProposals\ of $75$ runs and the
  hook records only 3 runs with an allowed attack call.

  The archived run and decision logs further decompose C's seven task failures.
  In two runs, preflight omitted the source-write capability required for the
  repair, so derivation withheld the patcher's legitimate write. In two other
  runs, the runner proposed the correct source edit but the workflow did not
  reroute it to a write-authorized patcher. The remaining three runs had passing
  tests at the end but reached the session timeout and are therefore counted as
  failures by the fixed scoring rule. Thus two failures directly expose the
  utility cost of a too-narrow ceiling; the other five reflect delegation or
  termination behavior rather than a missing task-wide permission.

  \paragraph{Why a Shared Task Policy Still Permits Attacks.}
  Because one policy is
  shared by every principal, it must grant the union of everything the task
  needs---reads, a source write, the test commands---just to let the repair
  finish, and that broad authority then reaches \emph{every} sub-agent, the
  runner that reads the poisoned output included. Removing a task-required
  permission would also prevent the legitimate step. \tool\ scopes authority
  per agent instead: the runner that reads the
  injection can run tests but cannot write source, so the attack is blocked,
  while the patcher still holds the write that completes the repair.

 \paragraph{Latency.}
Mean wall-clock time is \BZeroTime\,s for B0, \BOneTime\,s for B1,
\BTwoTime\,s for B2, and \BThreeTime\,s for C. For C, the median is
208\,s; five runs reached the 900\,s timeout, increasing the mean.
The decision logs provide a descriptive account of this additional work.
Among the 34 C runs containing an attack-targeting proposal, 27 contain
repeated attack-targeting calls. The measured latency therefore includes
additional model turns following refusals, rather than only the cost of
the authorization check. Derivation-related calls are also associated
with longer runs. These observations do not isolate the causal cost of
any individual component, since grant construction, sub-agent setup, and
other model interactions may also contribute.

  \paragraph{Interpreting the Cost.}
  The pipeline is fully autonomous and has no human approval path. Every
  denial returns control to the model, which may try another call. In an
  interactive deployment, a user could reject the first attack-driven request
  and end that retry sequence, reducing subsequent agent compute; the user could
  also approve a legitimate uncovered step. We did not measure human response
  time or approval choices, so this is a deployment implication rather than a
  measured latency result. Approval is disabled to keep the protocol fixed.

  \begin{table*}[t]
  \centering\small
  \begin{minipage}[t]{0.51\textwidth}
    \setlength{\tabcolsep}{3pt}
    \caption{Aggregate results per condition ($75$ runs each).
      \emph{Attack executed} (lower is better) is the headline defense metric,
      read from the filesystem; \emph{mean time} is wall-clock seconds per run.}
    \label{tab:case}
    \begin{tabular}{@{} l r r r r @{}}
    \toprule
    \textbf{Condition} & \textbf{Runs} & \makecell{\textbf{Task}\\\textbf{success}}
      & \makecell{\textbf{Attack}\\\textbf{exec.}} & \makecell{\textbf{Mean}\\\textbf{time\,(s)}} \\
    \midrule
    B0: ambient authority
      & $75$ & \BZeroTaskSuccess/75  & \BZeroLeaks/75  & \BZeroTime  \\
    B1: static global
      & $75$ & \BOneTaskSuccess/75   & \BOneLeaks/75   & \BOneTime   \\
    B2: task-specific global
      & $75$ & \BTwoTaskSuccess/75   & \BTwoLeaks/75   & \BTwoTime   \\
    C: \tool
      & $75$ & \BThreeTaskSuccess/75 & \BThreeLeaks/75 & \BThreeTime \\
    \bottomrule
    \end{tabular}
  \end{minipage}\hfill
  \begin{minipage}[t]{0.43\textwidth}
    \setlength{\tabcolsep}{5pt}
    \caption{Attack execution by injection surface for the strongest global
      baseline (B2) and \tool\ (C), with $15$ runs per cell.}
    \label{tab:surface}
    \begin{tabular}{@{} l r r @{}}
    \toprule
    \textbf{Surface} & \makecell{\textbf{B2 attack}\\\textbf{exec.}}
      & \makecell{\textbf{C attack}\\\textbf{exec.}} \\
    \midrule
    README         & $7/15$  & $0/15$ \\
    AGENTS.md      & $11/15$ & $3/15$ \\
    Skill file     & $7/15$  & $0/15$ \\
    Source comment & $4/15$  & $0/15$ \\
    Tool output    & $4/15$  & $0/15$ \\
    \bottomrule
    \end{tabular}
  \end{minipage}
  \end{table*}

  \section{Discussion}
  \label{sec:discussion}

  \paragraph{Policy Quality and Approval.}
  Enforcement checks the installed capabilities at every mediated call. Policy
  generation is a separate source of error: a scope that is too broad may admit
  an unwanted in-scope action, while one that is too narrow may block a required
  step. \tool\ does not widen a scope silently. A deployment can instead show
  the refusal to the user and record any approved capability as a fresh trusted
  \rulename{Mint}.

  \paragraph{How a Residual Injection Still Lands.}
  \tool's remaining executed attacks (Table~\ref{tab:case}) fall into two grant
  cases. (1) A grant can be wider than its subtask needs, such as permitting a
  directory write instead of one file. (2) Injected text can influence the
  model-generated subtask, causing the orchestrator or host-side grant
  completion to select an attacker-requested capability. Both cases require the
  action to lie within the preflight ceiling. Dispatch enforcement cannot
  correct a capability that the policy intentionally installed.

  \paragraph{Trusted Computing Base.}
  The TCB is exactly the host-side components named in
  Section~\ref{sec:threat}: the preflight isolation boundary, capability
  compiler, derivation code, per-principal stores, scope predicates, and
  dispatch hooks. The preflight LLM shapes how broad the initial policy is, but
  cannot place or alter capabilities once untrusted content is admitted---only
  host code can mint, narrow, or enforce them.

  \paragraph{Limitations and Threats to Validity.}
  The corpus contains five small Python repairs, one model, and three trials
  per task-surface-condition cell. It does not represent large refactors,
  dependency migrations, or long-running workflows, where policy omissions
  and approval frequency may be higher. The payloads are intentionally strong
  and the model may react differently to each carrier, so we report results by
  surface rather than assuming a uniform injection rate. Because the underlying
  calls are stochastic, individual cells vary from trial to trial; we therefore
  report aggregates over three trials and read the per-condition ordering, which
  is stable, rather than any single cell. Finally, the informative refusal
  message can affect recovery and task success, though it cannot turn an
  uncovered call into an allowed one.

  \section{Related Work}
  \label{sec:related}

  \paragraph{Prompt Injection and Model-Side Defenses.}
  Indirect prompt injection lets attacker-controlled external content hijack an
  LLM-integrated application~\cite{greshake2023not}. OWASP ranks it the leading LLM
  application risk~\cite{owasp2025llm01}. In the coding setting the threat
  is acute: Liu~\etal~\cite{liu2025youraishell} report attack success
  rates up to 84\% in production coding editors, and rule and skill files
  are a documented backdoor vector~\cite{pillar2025rules}. Model-side
  defenses such as Instruction Hierarchy~\cite{wallace2024instructionhierarchy}
  and PromptArmor~\cite{shi2025promptarmor} make the model more skeptical
  of untrusted text. Their decisions remain probabilistic. \tool\ instead
  checks installed permissions at the harness boundary and can be combined
  with model-side defenses.

  \paragraph{Harness-Level Defenses.}
  AgentDojo~\cite{debenedetti2024agentdojo} is a benchmark for agent prompt
  injection; Bhagwatkar~\etal~\cite{bhagwatkar2025firewalls} show that simple
  firewalls solve many cases in several benchmarks. Deployed coding agents also
  enforce permission boundaries through sandboxes, approval modes, and allowlist/denylist
  command patterns. \tool\ differs along two axes. Its command scopes match a
  parsed executable and argument vector, and a compound command is checked
  segment by segment rather than as one string; thus allowing \texttt{pytest}
  does not also admit \texttt{pytest; cmd} or \texttt{pytest \&\& cmd} unless
  \texttt{cmd} is itself covered, nor \texttt{\$(cmd)}, whose substituted
  argument matches no allowed invocation. Authority is also attenuated per
  agent.
  A task-wide global policy, even one generated specifically for the task,
  cannot permit a patcher to write a file while withholding that write from a
  runner. \rulename{Derive} can, and bounds the sub-agent by its delegation
  ceiling.

  Progent~\cite{shi2025progent} is the closest runtime-policy design: its DSL
  expresses symbolic rules over tool names and arguments, and a deterministic
  check mediates each call. An LLM generates its initial policy and may propose
  updates as execution evolves; narrowing updates are automatic, whereas
  expansions require approval. \tool\ instead freezes its task ceiling before
  repository content is read and additionally assigns a distinct attenuated
  store to each principal.

  Other systems govern agent execution at different points.
  CaMeL~\cite{debenedetti2025camel} uses capabilities in a dual-LLM planner
  with a custom interpreter. Fides~\cite{costa2025fides} tracks confidentiality and integrity
  labels over data flow; AgentArmor~\cite{wang2025agentarmor} reconstructs
  runtime traces into an IR and type-checks it; ToolGate~\cite{liu2026toolgate}
  attaches Hoare-style contracts to tool calls; and
  PACT~\cite{fan2026pact} treats injection as
  untrusted content determining an authority-bearing argument and enforces
  argument-level provenance contracts. \tool\ differs from all of these in
  where authority originates: it installs typed capability values from a
  trusted policy before untrusted content is read, then checks them at dispatch.
  This mechanism can be combined with data-flow tracking.

  \paragraph{Capability Security.}
  \tool's vocabulary is classical. Saltzer and
  Schroeder~\cite{saltzer1975protection} define a capability as an
  unforgeable authorization and elevate complete mediation and least
  privilege; Dennis and Van Horn~\cite{dennis1966programming} couple
  designation with authority, dissolving the confused-deputy
  problem~\cite{hardy1988confused}; and Miller and
  Shapiro~\cite{miller2003paradigm} distinguish authority from permission
  and show how attenuation follows from ordinary reference-passing.
  Bastion~\cite{omar2024bastion} maps an agent and task to an
  \mbox{F$^\star$} capability whose abstract monad exposes only permitted
  commands; a least-privilege model may restrict a baseline capability, and a
  verified typechecker checks the agent's monadic program. Odersky
  \etal~\cite{odersky2026tracked} instead make capabilities program values
  tracked by Scala's capture checker, allowing the type system to enforce
  capability safety and local purity, including confidentiality properties
  outside \tool's scope. \tool\ uses neither agent-held capability values nor
  static checking: it instantiates attenuation as structured scope predicates
  in host-side stores and enforces them at dispatch.

  \section{Conclusion}
  \label{sec:conclusion}

  \tool\ limits coding-agent tool calls with typed, host-side capabilities. A
  preflight step establishes a task ceiling before repository contents are
  read; delegation gives each sub-agent a narrower store; and a dispatch hook
  checks the issuing agent's store before executing a call. Our evaluation
  compares ambient authority, a static global policy, a task-specific global
  policy, and \tool\ across five tasks, five injection surfaces, and three
  trials per cell ($300$ runs). The baselines execute the injected effect
  in \BaseLeakLo--\BaseLeakHi/75 runs; \tool\ executes it in \BThreeLeaks/75
  and completes \BThreeTaskSuccess/75 repairs. The mechanism blocks a mediated
  call when the issuing agent lacks a matching installed capability. It does
  not prevent misuse of a granted capability or the transitive effects of an
  allowed command, which remain responsibilities of policy design and sandboxing.

\begin{acks}
This work was supported by the
\grantsponsor{GS501100001809}{National Natural Science Foundation of China
  (NSFC)}{https://doi.org/10.13039/501100001809}
under Grant No.~\grantnum{GS501100001809}{W2542035}.
\end{acks}

  \bibliography{references}
  \bibliographystyle{ACM-Reference-Format}
  \end{document}